\documentclass[aps, prb, twocolumn, amsmath, amssymb, showpacs, nofootinbib,superscriptaddress]{revtex4-2}

\usepackage[utf8]{inputenc}
\usepackage{graphicx,color,xcolor,times,amsmath}
\usepackage[pdftex, colorlinks=true, linkcolor=myblue, citecolor=myblue,urlcolor=myblue]{hyperref}

\DeclareOldFontCommand{\rm}{\normalfont\rmfamily}{\mathrm}

\definecolor{myblue}{rgb}{0,0,1}

\begin{document}
%
\title{Electronic heat tunneling between two metals beyond the WKB approximation}
\author{Mauricio G\'omez Viloria}
\email{mauricio.gomez-viloria@institutoptique.fr}
\affiliation{Université Paris-Saclay, Institut d'Optique Graduate School, CNRS, Laboratoire Charles Fabry, 91127, Palaiseau, France}
\author{Philippe Ben-Abdallah}
\affiliation{Université Paris-Saclay, Institut d'Optique Graduate School, CNRS, Laboratoire Charles Fabry, 91127, Palaiseau, France}
\author{Riccardo Messina}
\email{riccardo.messina@institutoptique.fr}
\affiliation{Université Paris-Saclay, Institut d'Optique Graduate School, CNRS, Laboratoire Charles Fabry, 91127, Palaiseau, France}
\date{\today}
\begin{abstract}
Two metals at different temperatures separated by large gaps exchange heat under the form of electromagnetic radiation. When the separation distance is reduced and they approach contact (nanometer and sub-nanometer gaps), electrons and phonons can tunnel between the bodies, competing and eventually going beyond the flux mediated by thermal photons. In this transition regime the accurate modeling of electronic current and heat flux is of major importance. Here we show that, in order to quantitatively model this transfer, a careful description of the tunneling barrier between two metals is needed and going beyond the traditional WKB approximation is also essential. We employ analytical and numerical approaches to model the electronic potential between two semi-infinite jellium planar substrates separated by a vacuum gap in order to calculate the electronic heat flow and compare it with its radiative counterpart described by near-field radiative heat transfer. We demonstrate that the results for heat flux and electronic current density are extremely sensitive to both the shape and height of the barrier, as well as the calculation scheme for the tunneling probability, with variations up to several orders of magnitude. Using the proximity force approximation, we also provide estimates for tip–plane geometries. The present work provides realistic models to describe the electronic heat flux, in the scanning-thermal-microscopy experiments.
\end{abstract}
\maketitle
\section{Introduction}

Two bodies at different temperatures separated by a vacuum gap can exchange heat through a variety of channels. At large separation distances this energy exchange is purely radiative and governed by the Stefan-Boltzmann law, setting an upper limit for this energy flux, reached only in the theoretical scenario of two blackbodies. When the separation distance becomes smaller than the thermal wavelength (of the order of 10\,$\mu$m at ambient temperature) we move into the regime of near-field radiative heat transfer (NFRHT) theory. In this domain, it is known that the radiative flux can exceed the Stefan-Boltzmann limit thanks to the contribution of evanescent (i.e. non-propagative) photons~\cite{Polder}. This strong flux amplification can reach several order of magnitude for materials supporting resonant surface modes of the electromagnetic field in the infrared, such as phonon-polaritons for polar materials~\cite{Joulain_rev,Volokitin_rev,RMP} or a continuum of hyperbolic modes~\cite{Biehs_prl}.

The physics at play becomes even richer when going to smaller distances, in the so-called extreme-near-field regime, at separation distances in the nanometer range and below. This distance regime has been recently probed by two experiments~\cite{reddy_17,kittel_17} reaching diverging conclusions, the former confirming theoretical predictions, the latter observing a strong flux amplification, to date unexplained. In the extreme-near-field regime, it has been shown that radiation can be influenced by nonlocal effects~\cite{fordweber,kittel_05,poc}, which could lead to new interesting phenomena, such as the existence of a radiative contribution stemming from non-optical modes between polar materials~\cite{arxiv_vibrational,arxiv_md}. It has also been argued that at sub-nanometer scales two new heat carriers contribute to energy transfer~\cite{Messina_arxiv,Francoeur1,guo22,nottingham23}. On the one hand, acoustic vibrations from a surface can have an influence on another surface due to molecular and electrostatic forces, leading to phonon tunneling~\cite{pendry16,pendry17,volokitin19,volokitin20,Francoeur2,guo22, nottingham23,arxiv_md}. On the other hand, when dealing with metals, electron tunneling is expected to significantly contribute and predicted to dominate close to contact. 

Besides the development of experimental setups probing heat flux in the extreme near field (for which the agreement with theory is often qualitative due to vibration, deformation and contamination~\cite{binnig_82}), the study of energy exchange at such short distance scales is also of remarkable importance due to recent and ongoing developments in nanofabrication and miniaturization. As a matter of fact, nanodevices need efficient thermal management techniques in order to be reliable, since slight temperature differences can drive significant uncontrolled amounts of heat. Motivated by these challenges, the study of the electronic contribution to energy exchange is of major importance. Moreover, the study of energy and heat flux by thermal electrons in the tunneling regime is of interest for the development of thermal transistors and thermal amplifiers~\cite{jchen16,jchen17}. ``Thermal" refers to electrons described by local equilibrium Fermi-Dirac statistics but below the work function. Electrons in the tail of the distributions are exchanged by tunneling if the barrier is thin, carrying both charge and heat. Under the influence of an electric potential bias, this can lead to the Nottingham effect where the electronic heat flux is large and non reciprocal~\cite{xu,nottingham23}, leading to the mutual heating of both electrodes. So far most studies comparing it with the radiative counterpart have modeled the effect of the barrier under a single model, including single-step potentials~\cite{Messina_arxiv} or under the influence of classical image forces~\cite{Francoeur1,guo22} which need to be regularized.

The study of the barrier height is of special importance for the study of surfaces in field emission and scanning tunneling microscopy (STM). For the latter, phenomenological or semiclassical formulas are derived to deduce the barrier height from the measured current~\cite{lang_barrier,simmons}. However these calculations need additional corrections depending on the electrodes, as deformations of the tip and the surface can lead to apparent barrier height and apparent gap distances, and thus apparent surfaces that differ from the expected results~\cite{teaguethesis,tersoff_86,solerSTM}. It also does not help that the current often varies exponentially with respect to various parameters, which limits the sensitivity to small feature changes~\cite{binnig_82}. Slight differences in chemical composition can lead to asymmetrical barriers which shift the conductance minima~\cite{barrier_asym}. Attractive forces can also appear near the surface producing a vibrating motion of the tip which in turn influences the measured barrier heights~\cite{teaguethesis}. The sensibility to the tip motion and deformability has led to the development of atomic force microscopy~\cite{AFM}. Another problem is contamination: even for clean surfaces and ultra-high vacuums, the work functions measured using these techniques can be lower than the expected value by some eV~\cite{chenSTM}. All of these issues imply that near-field scanning thermal microscopy (SThM)~\cite{kittel_05}, which adapts the equipment of STM and AFM to measure heat currents, suffers from the same problems in the presence of electronic heat transfer. Nevertheless, probing these heat exchanges could provide a secondary test for the barriers and interactions at extreme and near fields.

In this work we focus on providing a numerical bound to the electronic tunneling heat current by analyzing the effects of the modeling of the barrier in various extreme cases. The tunneling probability of electrons is calculated from a rigorous calculation based on the transfer-matrix method applied within a density functional approach to ideal jellium bodies as well as an analytic nonlocal Poisson equation under the Thomas-Fermi approximation~\cite{sidyakin,ilchenko80}. 
We also analyze the case of a parametrized phenomenological barrier given by a generalized Gaussian function. These approaches allow us to explore the influence of the height but also the shape of the barrier between two metallic electrodes.

This paper is organized as follows: The definitions of current density and heat flux are discussed in Section~\ref{sec:heatflux} for the case of thermal electron tunneling and NFRHT. In Section~\ref{sec:motivation}, we discuss the results of the classical potential under semiclassical approximations and illustrate the limitations of such an approach. Section~\ref{sec:barrierpot} is devoted to more realistic models for the electronic barrier potential between two metals and the calculation of the transmission probability. In Sec.~\ref{sec:results} we discuss the electronic heat flux in two different configurations, namely two metallic half spaces (plane--plane configuration) and a tip--plane configuration using the proximity force approximation (PFA) as shown in Fig.~\ref{fig:schema}. We finally conclude in Sec.~\ref{sec:conclusions}.

\section{Electronic current density and extreme-near-field heat flux}
\label{sec:heatflux}

\begin{figure}[t]
\includegraphics[width=\linewidth]{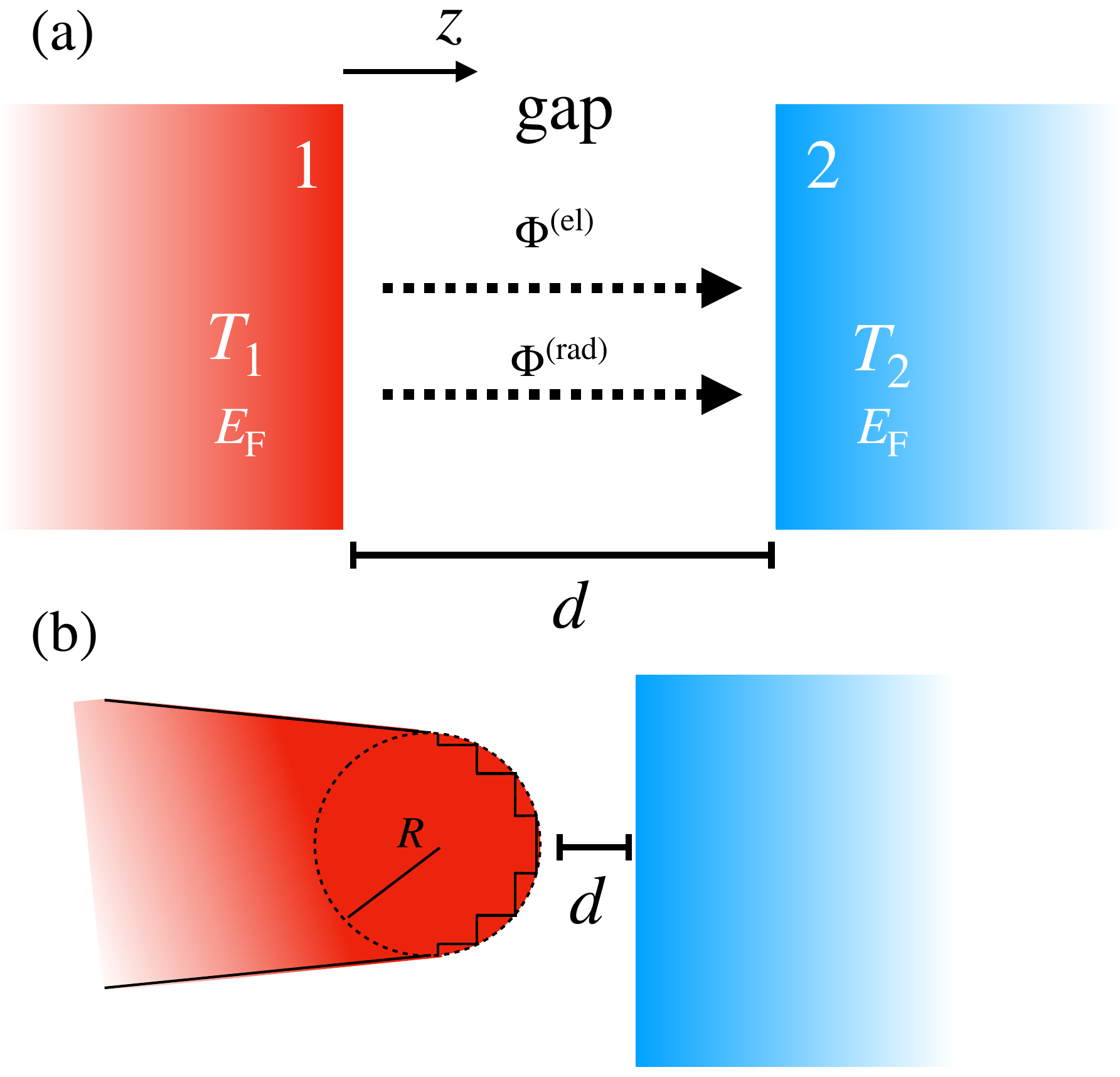}
\caption{\label{fig:schema}%
{Extreme near field heat fluxes between (a) two semi-infinite slabs of temperatures $T_1$ and $T_2$ made of the same metal with Fermi energy $E_\mathrm{F}$ separated by a vacuum gap of length $d$ and (b) between tip and sample with the same parameters. Near-field electromagnetic radiation (rad) and thermal electrons (el) can channel heat from body 1 to body 2. In (b), the tip is considered spherical with radius $R$, divided into infinitesimal disks for proximity force calculations. }
}
\end{figure}

Let us consider the system depicted in Fig.~\ref{fig:schema}(a), consisting of two metallic parallel planar substrates, separated by a vacuum gap of thickness $d$ along the $z$ direction.
They are assumed to be large enough along the $x$ and $y$ directions so that they can be considered infinitely extended. The two substrates are kept at two different temperatures $T_1$ and $T_2$ by two external thermostats and they are characterized by the same Fermi level $E_{\rm F}$. As explained above, even when separated by a vacuum gap, these bodies can exchange energy through the tunneling of different carriers. Here we are going to focus on electron tunneling, and take photon tunneling, i.e. radiative heat transfer, as a quantitative reference for comparison.

Even in the case of large work functions, electrons may escape the surface of a metal as a result of a temperature difference (thermionic emission) or in the presence of an externally applied electric field (field emission). The latter scenario is possible due to quantum tunneling, allowing for a non-zero transmission probability for electrons with classically insufficient energy to overcome the potential barrier in the region between the two substrates. More specifically, in the so-called extreme-near-field regime, i.e. when the barrier width is in the nanometer range and below, it is possible to induce a significant electron tunneling already for a small temperature difference and in the absence of a bias field. The tunneling current density in this configuration can be expressed as ~\cite{simmons} 
\begin{align}
	\label{eq:current}
	J=-\frac{e m}{2\pi^2 \hbar^3}&\int_{0}^{\infty}\mathrm{d}E_z\int_0^{\infty}\mathrm{d}E_{\perp}\\
	\nonumber&\times \Delta n_{\rm FD}(E,T_1,T_2,E_\mathrm{F}) \mathcal{T}^{\rm (el)}(E_z),
\end{align}
where $-e$ is the electron electric charge, $m$ its mass, $E=E_\perp+E_z$ its total kinetic energy decomposed in contributions stemming from velocities perpendicular and parallel to the surface, and 
\begin{equation}\label{eq:fermidirac}\begin{split}
	\Delta n_{\rm FD}(E,T_1,T_2,E_\mathrm{F})=n_{\rm FD}(E,T_1,E_\mathrm{F})-n_{\rm FD}(E,T_2,E_\mathrm{F}),
\end{split}
\end{equation}
$n_{\rm FD}(E,T_i,E_\mathrm{F})=1/[\exp([E-E_\mathrm{F}]/k_{\rm B}T_i)+1]$ being the Fermi-Dirac distribution that depends on both temperature $T_i$ and Fermi energy $E_\mathrm{F}$ associated with each medium. The key physical quantity appearing in Eq.~\eqref{eq:current} is the electronic transmission probability $\mathcal{T}^{\rm (el)}(E_z)$ for the electron to cross the gap, which due to the symmetry of the problem depends only on its kinetic energy $E_z$ perpendicular to the surface. The transmission probability has to be calculated by determining the transmission amplitude of a given electron crossing the gap in the presence of an electronic barrier $U(z)$ produced by image forces. The methods to calculate $\mathcal{T}^{\rm (el)}(E_z)$ are described in Sec.~\ref{sec:barrierpot}.

The net transfer of electrons between the two substrates is also at the origin of an energy flux (heat flow) $\Phi^{\rm (el)}$ which, as discussed in detail in \cite{nottingham23}, can in some configurations compete with and go beyond the photonic (radiative) heat flux $\Phi^{\rm (rad)}$. The total heat flux between the substrates is thus given by
\begin{equation}
\label{eq:totalflux}
\Phi=\Phi^{\rm (el)}+\Phi^{\rm (rad)}.
\end{equation}
We remark that in the extreme near-field regime, namely for gaps smaller than 1\,nm, one can also consider the possibility of phonon tunneling due to Van der Waals and electrostatic forces~\cite{pendry16,pendry17,volokitin19,volokitin20}, but this mechanism turns out to be a smaller contribution than the electronic one in the absence of bias~\cite{nottingham23} and will be neglected here. For the electronic contribution, the heat flux takes the form~\cite{xu,nottingham23}
\begin{align}
\label{eq:flux_el}
	\Phi^{(\mathrm{el})}&(T_1,T_2,d)=\frac{ m}{2\pi^2 \hbar^3}\int_{0}^{\infty}\mathrm{d}E_z\int_0^{\infty}\mathrm{d}E_{\perp}\\
	\nonumber&\times (E-E_\mathrm{F})\Delta n_{\rm FD}(E,T_1,T_2,E_\mathrm{F}) \mathcal{T}^{\rm (el)}(E_z),
\end{align}
where $(E-E_\mathrm{F})$ represents the energy contribution associated with each electron. Note that in the absence of bias voltage we make no distinction between the heat flow in the two directions (to and from cold and hot bodies), as in this case heat flows reciprocally in the usual thermodynamic way. This reciprocity does not always hold due to the Nottingham effect~\cite{xu,nottingham23}, which can lead for example to heating of both bodies in the presence of an applied bias voltage.

As stated above, radiative heat flux will be taken as a reference for comparison to electronic flux, since in the distance range considered here we can expect to find the transition separation distance below which electronic heat flux overcomes electromagnetic radiation~\cite{nottingham23}. The near-field radiative heat flux between the two bodies can be expressed as~\cite{Polder,pbajoulain,Age}
\begin{align}\label{eq:flux_rad}
	&\Phi^{\rm (rad)}(T_1,T_2,d)=\\
	&\nonumber\int_0^{\infty} \!\frac{\mathrm{d}\omega}{2\pi} \hbar \omega \Delta n_{\rm BE}(\omega,T_1,T_2) \int_0^\infty\!\frac{\mathrm{d} k}{2\pi}\; k\! \sum_{\alpha=\mathrm{s},\mathrm{p}}\mathcal{T}^{\rm (rad)}_\alpha (k,\omega, d),
\end{align}
where
\begin{equation}
	\Delta n_{\rm BE}(\omega,T_1,T_2)=n_{\rm BE}(\omega,T_1)-n_{\rm BE}(\omega,T_2),
\end{equation}
$n_{\rm BE}(\omega,T_i)=1/[\exp(\hbar \omega/k_{\rm B}T_i)-1]$ being the Bose--Einstein distribution, $k$ the parallel component of the wavevector, $\omega$ the angular frequency of each mode, and 
\begin{equation}\begin{split}
	\label{eq:transmission_rad}
	&\mathcal{T}^{\rm (rad)}_{\alpha}(k,\omega,d)\\
	&\,=\begin{cases} 
		\displaystyle\frac{(1-|r_{\alpha}|^2)^2}{|1-r_{\alpha}^2\exp(2\mathrm i k_z d )|^2}, & k<\omega/c,\\\vspace{-0.3cm}\\
		\displaystyle\frac{4\,(\mathrm{Im}\,r_{\alpha})^2\exp(-2\,\mathrm{Im}\,k_z d)}{|1-r^2_{\alpha}\exp(-2\,\mathrm{Im}\,k_z d )|^2}, & k\geq \omega/c,
	\end{cases}
\end{split}\end{equation}
the radiative transmission probability. The transmission probability is separated in terms of the two polarizations, given by the transverse electric ($\alpha=\mathrm s$) and transverse magnetic $(\alpha=\mathrm{p})$ contributions, where $k_z=\sqrt{(\omega/c)^2-k^2}$.
The integral in Eq.~\eqref{eq:flux_rad} is carried out over all values of $k$, including the contribution of propagative ($k<\omega/c$) and evanescent ($k>\omega/c$) waves. The latter dominate for distances below the thermal wavelength, in the micrometer range at ambient temperature. The reflection coefficients in \eqref{eq:flux_rad} are given by Fresnel's formulas,
\begin{equation}
\label{eq:r_s}
	r_{\mathrm s}(k,\omega)=\frac{k_z-k_{\mathrm m, z}}{k_z+k_{\mathrm m, z}},\quad r_{\mathrm p}(k,\omega)=\frac{\epsilon(\omega) k_z-k_{\mathrm m, z}}{\epsilon(\omega) k_z+k_{\mathrm m, z}},
\end{equation}
where $k_{\mathrm m,z}=\sqrt{ (\omega/c)^2 \epsilon(\omega)-k^2}$ is the $z$ component of the wavevector inside the media. In this paper, we employ a local description of the dielectric susceptibility given by Drude's model, as
\begin{equation}
	\epsilon(\omega)=\epsilon_{\infty}- \frac{\omega_{\rm pl}^2}{\omega(\omega+\mathrm i \Gamma)},
\end{equation}
where $\omega_{\rm pl}$ is the plasma frequency of the metal, $\Gamma$ is the damping coefficient and $\epsilon_{\infty}$ is the high frequency value~\cite{constants}. Here we neglect the nonlocal radiative effects that appear in the extreme near-field regime~\cite{fordweber,kittel_05,poc}, as this modification is negligible when compared to electronic tunneling~\cite{nottingham23}.

\section{The standard modeling of electron tunneling}
\label{sec:motivation}
The evaluation of the electronic transmission probability in Eq.~\eqref{eq:current} and Eq.~\eqref{eq:flux_el} depends on the calculation of the electronic barrier potential.
For a charge between two metallic plates, this potential is classically calculated using the image method~\cite{simmons}, given by the classical image potential,
\begin{equation}
\label{eq:image_potential}
	U_{\rm cl}(z)=W_0+E_{\rm F}+ \frac{e^2}{16\pi \epsilon_0 d } \left[\Psi(z/d)+\Psi(1-z/d)+2\gamma \right],
\end{equation}
only defined between $z=0$ and $z=d$, where $W_0$ is a vertical shift, $\epsilon_0$ is the vaccum permittivity, $\Psi(z)$ is the digamma function and $\gamma$ is the Euler-Mascheroni constant. The last term to the right of Eq.~\eqref{eq:image_potential} is known as the image potential or image force and has the effect of rounding the edges of a square barrier of height $E_{\rm F}+W_0$. The height of the barrier is reduced by the image potential, so $W_0$ is not the true work function. The image potential also reduces with the gap $d$. However, this expression \eqref{eq:image_potential} can be unphysical. 
Due to its divergences at the boundaries, the potential should be impenetrable~\cite{impenetrable} and semiclassical calculations of the transmission, which require smooth potentials, may not be valid. The presence of electronic interactions leads to a barrier that is actually smooth and penetrates into the metal. To avoid this issue, it is often suggested to redefine the image planes and translate the potential inside the metal by a few angstroms.
 
In order to compare with different barriers, we are interested in studying carefully the influence of height and shape of the barrier in a more general scenario. Thereby, we introduce a parametrized barrier described by a symmetric generalized Gaussian (GG) distribution~\cite{generalizedgaussian}, defined as
\begin{equation}\label{eq:UGG}
	U^{\rm (GG)}(\alpha,\beta,u_0;z)=u_0E_\mathrm{F} \exp\left(-\left|\frac{\frac{z}{d}-\frac{1}{2}}{\alpha}\right|^\beta\right),
\end{equation}
where the barrier is centered at $d/2$ and has three main parameters: the normalized scale parameter $\alpha$ quantifying the penetration of the potential inside the metal, the shape parameter $\beta$ which controls its peakedness and the barrier height $u_0$ (in units of the Fermi energy $E_{\rm F}$). The latter is connected to the work function $W$ by the simple relation $W=E_{\rm F}(u_0-1)$. This function also allows for long tail distributions, but in order to ensure that $U^{\rm (GG)}(\alpha,\beta,u_0;z)\to 0$ as $z\to \pm\infty$, we add the restriction $\beta\geq 1$.
Equation \eqref{eq:UGG} describes a standard Gaussian distribution for $\beta=2$ and approaches the square barrier as $\beta\to \infty$. The GG barrier allows to obtain results for general barrier shapes and heights. The generalized Gaussian potential has the advantage that it is also defined inside the metal.

\begin{figure}[t]
\includegraphics[width=\linewidth]{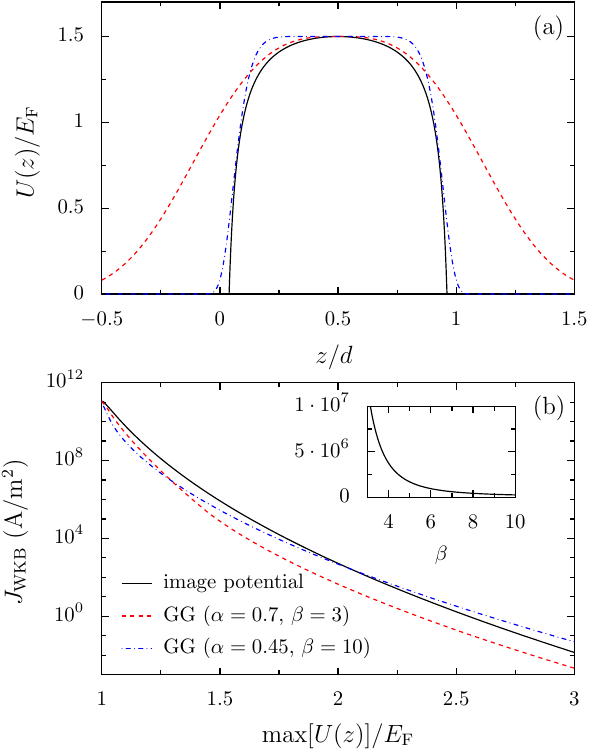}
\caption{\label{fig:WKBonly}%
(a) Shape of three electronic potential barriers of height of 1.5 $E_{\rm F}$: image potential (black solid line), GG barrier with $(\alpha,\beta)=(0.7,3)$ (red dashed line), GG barrier with $(\alpha,\beta)=(0.45,10)$ (blue dash-dotted line). (b) Current density calculated within WKB approximation for the same potential barriers as (a) as a function of the barrier height for a gap of $d=1\;$nm. Inset: current density as a function of the shape factor $\beta$ for a GG barrier of $\alpha=0.45$ and height $u_0=1.5$.
}
\end{figure}

In Fig.~\ref{fig:WKBonly}(a), we illustrate the image potential (black solid line) and two GG parametrizations corresponding to two different barriers with the same height and different shape that penetrate inside the metal ($z/d<0$ and $z/d>1$).

For a given parametrization of $U^{\rm (GG)}(\alpha,\beta,u_0;z)$, we can now solve for the transmission of an electron with energy $E_z$ inside the metal. However the transmission probability has only a few analytical solutions tied to specific electronic barrier shapes. In practical applications, the barrier height is often estimated qualitatively by using semiclassical approximations like that of the one-dimensional Wentzel--Kramers--Brillouin (WKB) method~\cite{WKB}, where the transmission is given by 
\begin{align}
\label{eq:WKB}
	\mathcal{T}^{\rm (el)}_{\rm WKB}&(E_z)=\\
	\nonumber &\exp\left(-\frac{2\sqrt{2m}}{\hbar}\int_{z_1}^{z_2} \mathrm d z\;\sqrt{U(z, V_{\rm b})-E_z}\right),
\end{align}
where the integration is usually carried out between the zeros of the integrand, $z_1$ and $z_2$, in the region where the electronic barrier height $U(z)$ is larger than the energy $E_z$. However the WKB approximation is drastic and should be avoided for extremely small gaps. Even if it is often the preferred technique for the calculation of the transmission of one-dimensional barriers, this approximation is not valid in the presence of abrupt potentials and in principle should be avoided when using the classical image potential~\eqref{eq:image_potential}, which can be proved to be impenetrable~\cite{impenetrable}.

In Fig.~\ref{fig:WKBonly}(b) we show the current density \eqref{eq:current} for a gap of $d=1\;$nm calculated using the WKB approximation, for the potentials defined in Fig.~\ref{fig:WKBonly}(a), where the integration in Eq.~\eqref{eq:WKB} is carried between $z_1=0$ and $z_2=d$.
As expected the current density can decrease by various orders of magnitude as a function of the height of the barrier. However it is also interesting to observe the sensitivity to the shape of the potential. For the broadest GG barrier ($\alpha=0.7$, red dashed line) the difference with the current density of the image potential can reach more than an order of magnitude depending on the height. For the thinner GG barrier ($\alpha=0.45$, blue dash-dotted line), the barrier is more similar to the image potential, but its relative difference with current density of the image potential can vary non-monotonically. 
For a given GG barrier, increasing the shape factor $\beta$ of the barrier from a peaked distribution to a square potential can induce a reduction of one order of magnitude or more of the current density even if the barrier height is kept at the same value [see, inset of Fig.~\ref{fig:WKBonly}(b)].

It is clear that just estimating the height of the barrier does not suffice to provide a quantitative calculation that would match experimental results. 
A given experimental data point of the tunneling current can be reproduced by any series of slightly similar potentials. It is possible to fit the data by adjusting the height $W$ and images planes of the classical image potential~\eqref{eq:image_potential} or by choosing a smooth potential by changing the height, shape and penetration depth. Also by postulating variable parameters one is able to fit any dependence of the current density with distance. This arbitrary choice makes it hard to understand what would be the next correction to the standard theory of electronic tunneling, as any divergence from experimental values can be identified as an unusual work function if one does not account for the changes in barrier shape or lack of sensitivity due to the use of the WKB approximation.
These problems motivate the inquiry to understand how much discrepancy there can be between less arbitrary theoretical models. When driven by a bias voltage, the classical image potential along with WKB approximation can be enough to broadly estimate the current density for gaps of a few nm~\cite{simmons} but would remain a very rough approximations for smaller gaps where the shape of the barrier changes and WKB is not sensitive at all to the penetration of the potential inside the barrier. The fact that in some experiments the work function of metals is mysteriously low~\cite{reddy_17,binnig_82} might depend on these approximations. 

\section{Beyond the standard approach}
\label{sec:barrierpot}

In the this section, we describe two different models that go beyond the classical image potential \eqref{eq:image_potential}, one based on a many-body calculation using density functional theory and another based on a nonlocal electrostatic solution of Poisson equation. With the goal of obtaining quantitative results for the heat flux from these models, we will also drop the WKB approximation altogether and replace it with a more precise numerical calculation of the transmission coefficient based on $S$-matrices.

\subsection{Local density approximation for jellium}
\label{sec:LDA}

Due to the unphysical nature of the classical images potential~\eqref{eq:image_potential}, a realistic calculation of the electronic barrier for a metal requires a many-body treatment of electronic interactions. In this approach, we model the electron gas inside the metallic bodies as a jellium (interacting electron cloud over a positive ionic background) and the effective potential that the electronic cloud exerts on a single probe electron is recovered. The jellium model has the advantage that it only depends on a single parameter, the Wignez-Seitz radius, which makes it very practical for the study of metals. It also allows us to clearly keep defined edges of the metal gap using a sharp ionic background. According to the Hohenberg--Kohn theorem~\cite{HKtheorem}, the total many-body energy can be written uniquely in terms of the electronic density $n$ as
\begin{align}\label{eq:dftenergy}
	\mathcal{E}[n]&=K[n]+\int \mathrm d^3 \mathbf r\; U_{\rm ext}(\mathbf r )n(\mathbf r) \\
	&+\frac{e^2}{8\pi \epsilon_0}\int \mathrm d^3\mathbf r\int\mathrm d^3\mathbf r'\; \frac{n(\mathbf r)n(\mathbf r')}{|\mathbf r-\mathbf r'|}+\mathcal{E}_{\rm xc}[n],
\end{align}
where $K[n]$ is the kinetic energy of a non-interacting electron gas, the second term represents the interaction with an external potential $U_{\rm ext}(\mathbf r)$, the third term is the electron-electron interaction and $\mathcal{E}_{\rm xc}$ represent the exchange-correlation contribution~\cite{lang70}. For the problem at hand, we are interested in the effective potential
\begin{equation}
	U(\mathbf r)=U_{\rm ext}(\mathbf r) + \frac{e^2}{4\pi\epsilon_0}\int\mathrm d^3\mathbf r'\; \frac{n(\mathbf r')}{|\mathbf r-\mathbf r'|}+\frac{\partial \mathcal{E}_{\rm xc}[n]}{\partial n(\mathbf r)},
\end{equation}
acting on single electron and due to the surrounding electrons.

By choosing a form of $\mathcal{E}_{\rm xc}[n]$, we can then solve the Kohn-Sham equations,
\begin{equation}\label{eq:kohnsham}
	\left[-\frac{\hbar^2}{2m}\nabla^2+U(\mathbf r)\right]\psi_j(\mathbf r)=E_j\psi_j(\mathbf r )
\end{equation}
for $j=1,2,\cdots,N$, for a system of $N$ electrons in a given volume, to obtain back the electronic density $n(\mathbf r)=\sum_{j=1}^N |\psi_j(\mathbf r)|^2$. By iterating over this self-consistent system of equations, Eqs. \eqref{eq:dftenergy} and \eqref{eq:kohnsham}, one can obtain a realistic approximation of the electronic density and the effective potential in the gap and inside the metal.

The exchange-correlation term is not known and requires to be treated under certain approximations. For this manuscript, we will restrict our calculations to the local-density approximation (LDA) \cite{kohnsham65}, which assumes an exchange-correlation function of the form $\mathcal{E}^{\rm (LDA)}_{\rm xc}=\int \mathrm d^3 \mathbf r\;n(\mathbf r) \varepsilon_{\rm xc} [n(\mathbf r)] $ where $\varepsilon_{\rm xc}=\varepsilon_{\rm x}+\varepsilon_{\rm c}$ is the exchange-correlation energy per electron for jellium. This term can be divided into two terms: a Fock exchange term $\varepsilon_{\rm x}$ that can be written analytically for a homogeneous electron gas, and a correlation term $\varepsilon_{\rm c}$ that is often obtained by quantum Monte-Carlo methods. In the present work, we implement the exchange-correlation potential using the Perdew and Yang approach~\cite{perdewyang}. 
For two semi-infinite jellium slabs with perfectly flat surfaces separated by a vacuum gap, this construction leads to an effective LDA barrier $U^{\rm (LDA)}(\mathbf r)=U^{\rm (LDA)}(z)$ that we calculate numerically using the GPAW toolkit~\cite{gpaw1,gpaw2} (see Appendix \ref{app:DFT} for technical details).

\subsection{Thomas–Fermi approximation for the nonlocal Poisson equation}
\label{sec:TFA}

The LDA approximation for jellium neglects the crystal structure of the material making it inadequate for the description of surface effects in metals. Moreover, it can provide low values for the work function of metals~\cite{chenSTM}. For that reason, we propose another model that would be closer to the semiclassical calculation, but where the work function is not an input of the model like in the classical image potential \eqref{eq:image_potential}. Inspired by previous results~\cite{nottingham23,constants}, we reintroduce here an additional barrier based on the analytical solution to the nonlocal Poisson equation, ~\cite{sidyakin,ilchenko80,nottingham23}, given by
\begin{equation}
\label{eq:nonlocalPoisson}
\begin{split}
	\left(\frac{\partial^2 }{\partial z^2}-k^2\right) G(k;z,z') - &\int\mathrm{d} z^{\prime \prime} \Pi(k;,z,z')G(q; z^{\prime\prime},z')\\
	&=\delta(z-z'),
\end{split}\end{equation}
where $\delta(z)$ is the Dirac delta distribution, $G(k;z,z')$ is the Green function and $\Pi(k;z,z')$ is the polarization operator. Using the specular reflection approximation, we can write the polarization operator as
\begin{equation}
	\Pi(k;z,z')=\begin{cases}
	\Pi_1(k;z-z')+\Pi_1(k;z+z'),\\
	\hspace{4cm} z,z'\leq 0,\\
	\Pi_2(k;z-z')+\Pi_2(k;z+z'),\\
	\hspace{4cm} z,z'\geq d,\\
	\Pi_{\rm gap}(k;z-z')+\Pi_{\rm gap}(k;z+z'),\\
	\hspace{4cm} 0 < z,z' <d,
	\end{cases}
\end{equation} 
where
\begin{equation}
	\Pi_b(k;z\mp z')=\int_{-\infty}^\infty \frac{\mathrm{d}q_z}{2\pi} K^2[\epsilon_b(K)-1]\exp(\mathrm i q_z[z\mp z']),
\end{equation}
$K^2=k^2+q_z^2$, and $\epsilon_{b}(K)$ is the dielectric function of each region $b=1,2,\mathrm{gap}$. For simplicity, we would use the long wavelength Thomas-Fermi approximation (TFA) for the dielectric function \cite{ashcroft} inside the metal, given by
\begin{equation}
\label{eq:epsilon_TF}
	\epsilon_{\rm TF}(K)=\epsilon_{1,2}(K)=1+\frac{ k_{\rm TF}^2}{K^2},
\end{equation}
where $k_{\rm TF}=\sqrt{e^2 m_{\rm e} k_{\rm F}/\pi^2 \hbar^2\epsilon_0}$ is the inverse of the Thomas--Fermi screening length \cite{ashcroft,constants} and it is the only input parameter for the calculation. By solving for $G(k;z,z')$ in Eq.~\eqref{eq:nonlocalPoisson} \cite{ilchenko80,nottingham23} we can recover the electronic potential for a single electron by calculating 
\begin{equation}
	U^{\rm (TFA)}(z)=\frac{e^2}{4\pi \epsilon_0} \left\{\frac{k_{\rm TF}}{2}- \int_0^\infty \mathrm{d}k\,k\left[G(k;z)+\frac{1}{2k}\right]\right\}.
\end{equation}
where the constant $e^2k_{\rm TF}/8\pi\epsilon_0$ is introduced to set the bottom of the band equal to 0. The Thomas-Fermi approximation considered the first valid approximation beyond the classical image potential used to reproduce screening effects, but does not reproduce other quantum phenomena like the Friedel oscillations of the electronic density. The TFA barrier reproduces the classical potential \eqref{eq:image_potential} for ideal metals [$\epsilon_{1,2}(K)\to \infty$].

\subsection{Comparison}

To go beyond the classical image potential \eqref{eq:image_potential}, we have introduced two different models that go beyond the classical assumptions, the TFA potential of Sec.~\ref{sec:TFA} which introduces screening effects and the LDA approach of Sec.~\ref{sec:LDA}, which treats the quantum many-body problem and adds the effects of exchange and correlation potentials. On the one hand, the effective TFA barrier $U^{\rm (TFA)}(z)$ is simpler to implement, but is known to overestimate the size of the barrier. On the other hand, $U^{\rm (LDA)}(z)$ for jellium is a more complete treatment but requires numerical effort and it is well known to underestimate the work function of metals~\cite{chenSTM}. For these reasons, the LDA and the nonlocal TFA barrier serve to set two limits for the height and shape of the effective barrier.

\begin{figure*}[t]
	\includegraphics[width=\linewidth]{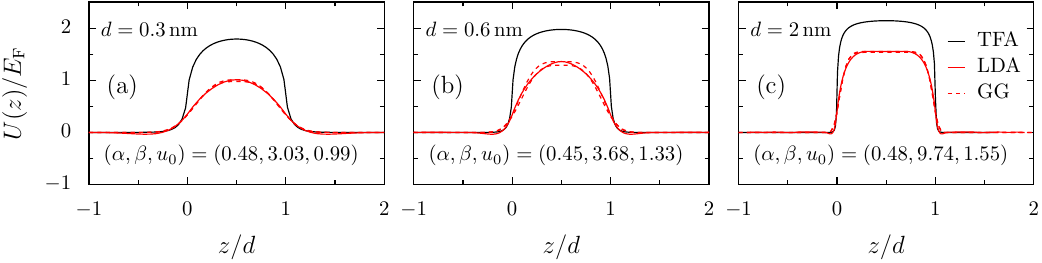}
	\caption{\label{fig:barriershape}%
		{Effective electronic barrier potential between two semi-infinite slabs of the same metal as a function of the horizontal coordinate $z$, for three different distances $d$, calculated using LDA and TFA methods (described in the text), along with the fits of the LDA barrier using a parametrized generalized Gaussian (GG) function with parameters $\alpha,\beta,u_0$ (see text for details).}}
\end{figure*}

Additionally, the parametrized GG barrier from Eq.~\eqref{eq:UGG} is simple enough to allow us to fit both TFA and LDA barriers, and to compare the heat flux and current related to these models. 
The LDA and TFA effective electronic barriers are shown in Fig.~\ref{fig:barriershape} for three distances $d=$0.3, 0.6 and 2\,nm.
We remark that, as anticipated, the LDA curve (in red) is always smaller than the TFA curve (in black) coming from nonlocal Poisson equation.
Contrary to the predictions of the classical image potential, the LDA and TFA barriers are not divergent and penetrate into the metal.
In order to fit the LDA curves using the GG function from Eq.~\eqref{eq:UGG}, we can either fit with respect to the three parameters $\alpha,\beta$ and $u_0$, or fix the value of $u_0$ as equal to the barrier maximum divided by $E_{\rm F}$ and then fit with respect to $\alpha$ and $\beta$. As shown in Fig.~\ref{fig:barriershape}, the former (latter) choice results in an underestimated (overestimated) function. This procedure allows us to define an average value for $\alpha,\beta$ and $u_0$ (reported in Fig.~\ref{fig:barriershape} for each distance) along with an error bar for the three parameters.

For large $d$, both the LDA and TFA barriers tend to a square step potential (corresponding to a large $\beta$). For $d<1$~nm, the barrier maxima of LDA and TFA potentials decrease with distance. The LDA barrier can take some negative values, but this effect is not as pronounced as in the Friedel oscillations of the electronic density due to the exchange-correlation contribution~\cite{lang70}.

\begin{figure}[pth]
\includegraphics[width=0.9\linewidth]{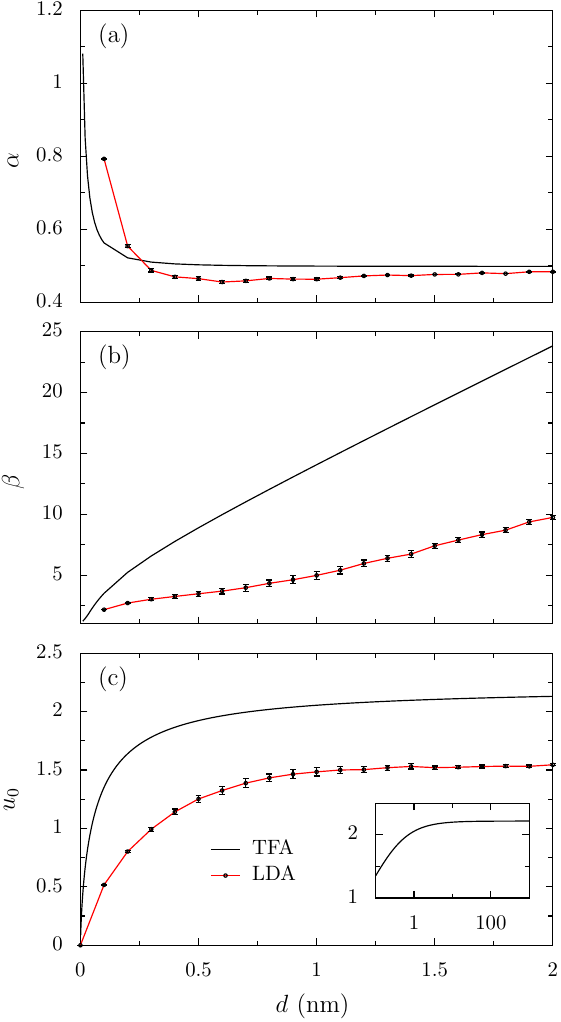}
\caption{\label{fig:alphabetau}%
{(a) Scale parameter $\alpha$, (b) shape parameter $\beta$ and (c) relative barrier height $u_0$ of GG potential as a function of the gap size $d$, when fitted to the TFA (black) and LDA (red) effective potentials. Error bars indicate fit errors. Inset: TFA $u_0$ as a function of distance (nm) for larger distances.}
}
\end{figure}

The dependence of the GG fitting parameters with respect to the distance is shown in Fig.~\ref{fig:alphabetau}. 
Close to contact, the LDA barrier gets significantly more Gaussian $\beta\lesssim 3$ or even close to a Laplace distribution $1<\beta< 2$.
The LDA value for the barrier height $u_0$ (in relative units with respect to $E_{\rm F}$) is below the Fermi level for gap distances smaller than about 3~\AA{}, which could be interpreted as contact since most electrons are no longer tunneling and are actually delocalized between the two bodies, meaning that in this case the work function is not properly defined. We have verified that the value of $u_0$ for the LDA barrier for gaps larger than 1 nm already coincides with the expected theoretical value for the jellium work function for gold~\cite{chenSTM} for a single surface, which is lower than the experimental value by 3 to 4\,eV. Conversely, the asymptotic value of $u_0$ for the TFA barrier overestimates the barrier height by the same amount and shows a much slower convergence (see the inset of Fig.~\ref{fig:alphabetau}). In this scenario an estimate for the work function can be obtained by employing again the relation discussed above $W=E_{\rm F}(u_0-1)$. The sharp decrease in the height of the TFA barrier close to contact is comparable to the apparent barrier height that is found in experiments, where the height remains constant when reducing the distance up to a couple of \AA{}~\cite{chenSTM}. The scale factor $\alpha$ does not vary much but close to contact diverges indicating delocalization and larger penetration of the effective potential into the metal. 

\subsection{Transmission probability using the $S$-matrix method}

As we want to quantitatively account for the barrier shape, we work with a quantum mechanical description of the barrier alongside a more accurate algorithm based on the scattering $S$-matrix algorithm from multilayered optics~\cite{smatrix1} to calculate the electronic transmission probability. This method accounts for oscillations of the transmission at large energies and the shape of the barrier inside the metal. 
The $S$-matrix algorithm provides the same results as the transfer-matrix method \cite{Tmatrixmethod}, which consists of dividing the barrier in differential slices and multiplying the transfer matrices of each slab.

Instead of using transfer matrices we calculate the scattering matrix of the $i$-th slice and multiply them together in sequence using the Redheffer star product~\cite{redheffer}. In the end, we recover the total $S$-matrix of the barrier for a given electron energy from which the transmission probability can be extracted.
The $S$-matrices are preferred here over the transfer matrix method for their numerical stability for large gaps~\cite{smatrix1}.

\begin{figure}[t]
\includegraphics[width=\linewidth]{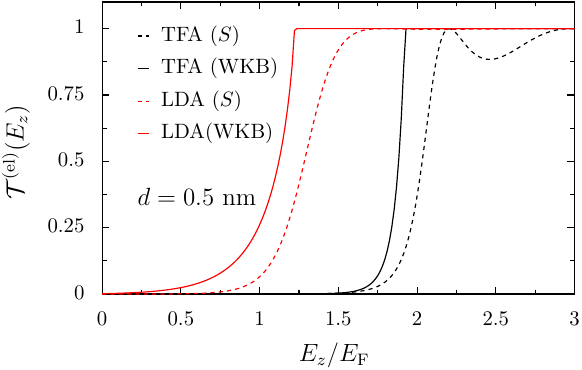}
\caption{\label{fig:WKBtrans}%
Electronic transmission probability as a function of kinetic energy $E_Z$ (in units of $E_{\rm F}$) for a gap of $d=5$~\AA{}. Two barriers are presented TFA and LDA, under two different calculation methods: $S$-matrix algorithm and WKB method. 
}
\end{figure}

The electronic transmission probability, used in the equations of the current density \eqref{eq:current} and of the heat flux~\eqref{eq:flux_el}, is plotted in Fig.~\ref{fig:WKBtrans} for an intermediate gap distance $d$ of 5~\AA{}.
It can be seen that the two methods barely agree qualitatively, as the WKB method of Eq.~\eqref{eq:WKB} is well known to neglect the oscillation of the transmission for electrons with energy higher than the barrier height as seen in the case of the TFA potential, whereas for the LDA barrier the oscillations of the tranmission are less drastic due to the smoothness of the potential (small $\beta$). The WKB transmission rises much more rapidly than in the $S$-matrix calculation which will lead to an overestimate of both current and flux. This rapid increase is less drastic for the TFA case, which is expected as the WKB approximation will approach the one calculated with the $S$-matrix algorithm for larger distances and barrier heights. We clarify that aside from the figures where it is labeled as such, we do not employ the WKB method in the reported calculations anywhere else in this manuscript.

\section {Results}
\label{sec:results}
\subsection{Plane--plane configuration}

\begin{figure*}[th]
\includegraphics[width=\linewidth]{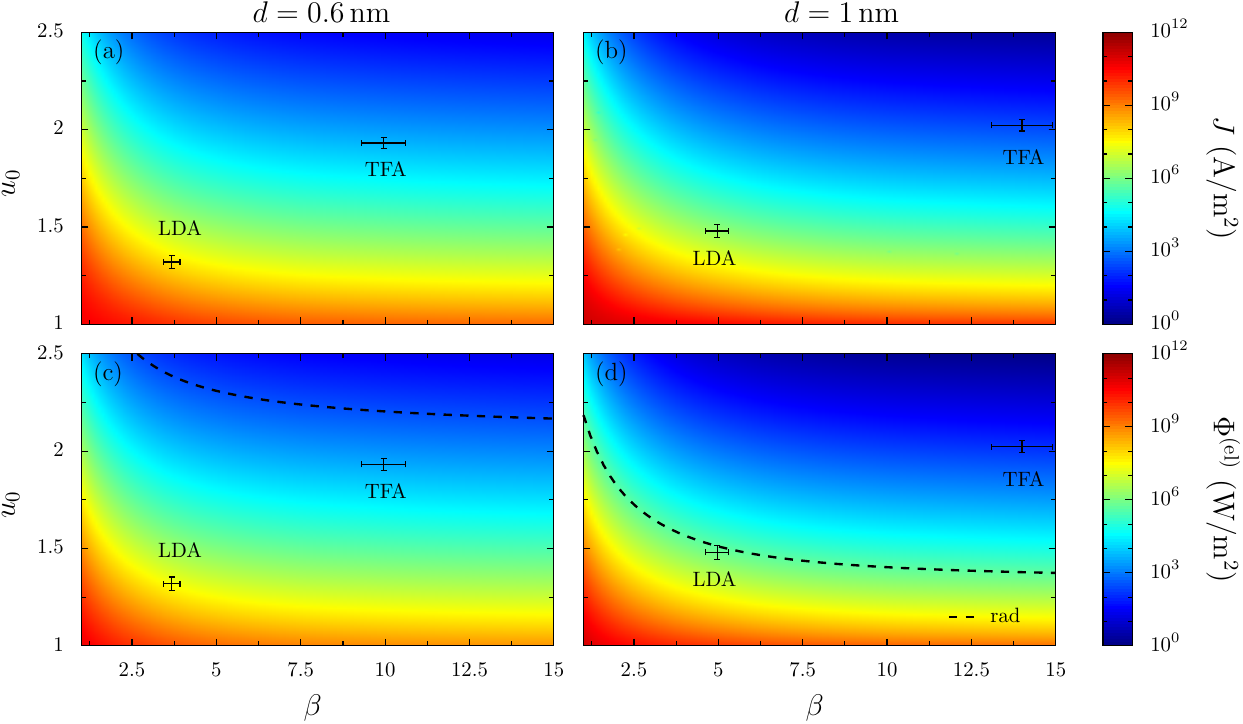}
\caption{\label{fig:colormaps}%
Density plot of the the current density $J$ (color axis, upper panels) and electronic heat flux (color axis, lower panels) between two semi-infinite metals with temperatures $T_1=400$ K and $T_2=300$ K, described by GG barrier and for gap distances of 0.6 (left panels) and 1~nm (right panels), as a function of shape factor $\beta$ and $u_0$ (for $\alpha=0.48$). Dashed line indicates the value of the near field radiative heat flux for the given distance. The corresponding mean $\beta$ and $u_0$ to fit the TFA and LDA barriers are indicated by points in the figure, along with associated error bars.}
\end{figure*}

In this section, we discuss electronic tunneling in the case of the plane--plane configuration as illustrated in Fig~\ref{fig:schema}(a). For all figures, we consider temperatures $T_1=400$~K and $T_2=300$~K.
The current density and electronic heat flux (color axis) as a function of the shape factor $\beta$ and relative barrier height $u_0$ are shown in Fig~\ref{fig:colormaps}, calculated using Eqs.~\eqref{eq:current} and \eqref{eq:flux_el} for a GG barrier for two different distances.
Both current and heat flux are presented in logarithmic scale in order to show the strong discrepancies that can be obtained by slight changes in the height but also in the shape $\beta$ of the barrier. The specific cases of LDA and TFA are marked by points in Fig.~\ref{fig:colormaps}, with associated error bars. 
Not only do the positions of these points show the extreme sensitivity of both current and heat flux to the choice of the barrier shape, but in the configuration of Fig.~\ref{fig:colormaps}(d) ($d=1\,$nm) LDA and TFA even lead to opposite conclusions on the comparison between electronic and photonic flux. Note that our calculations do not include the surface roughness which is considered to reduce the height of the barrier~\cite{binnigSTM}. These results highlight the importance of understanding the realistic shape of the barrier as it can lead to difference in the order of magnitude of the current and electronic heat flux. We also confirm that the electronic heat flux can overcome the radiative heat flux at least for distances smaller than 1\,nm, independently of the model.

\begin{figure}[b]
\includegraphics[width=\linewidth]{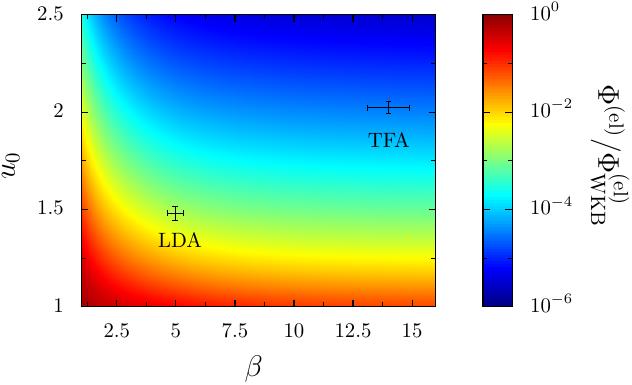}
\caption{\label{fig:WKB}%
Density plot of the ratio between the electronic heat flux $\Phi^{\rm (el)}$ calculated using the $S$-matrix algorithm and $\Phi^{\rm (el)}_{\rm WKB}$ from the WKB method (color scale) as a function of the relative height $u_0$ and the shape $\beta$ for a 1\,nm GG barrier. The corresponding mean $\beta$ and $u_0$ to fit the TFA and LDA barriers are indicated by points with associated error bars.
}
\end{figure}

After discussing the impact of the barrier shape, we focus on the method employed to calculate the tunneling probability. To this aim we compare in Fig.~\ref{fig:WKB} the electronic flux calculated with the $S$-matrix algorithm and the WKB method.
The figure clearly shows that the difference in the calculation method can lead to disagreements of several orders of magnitude depending on both the barrier height $u_0$ and shape $\beta$. In experiments, one could be able to correct the WKB estimation of the barrier by comparing it with a more precise transmission calculations. However if the shape of the barrier is not properly taken into account one still risks to underestimate the barrier height. Both precise numerical and WKB methods would seem only to agree in extreme cases where the barrier is shallow or very peaked (low $\beta$). 

\begin{figure}[t]
\includegraphics[width=\linewidth]{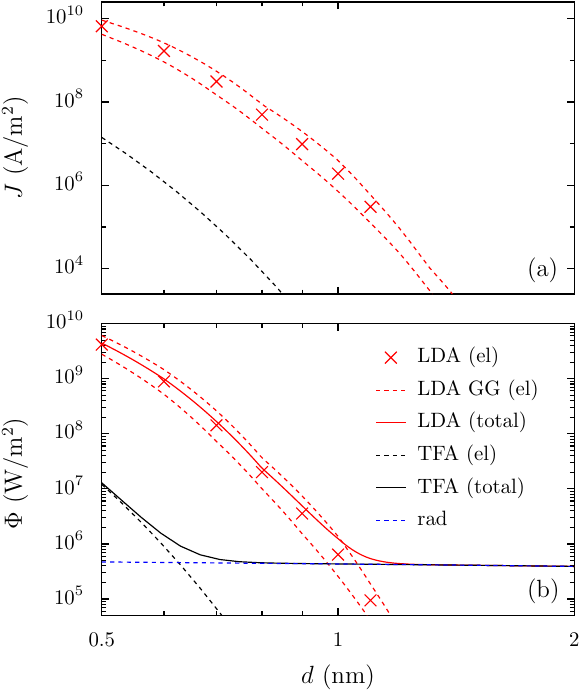}
\caption{\label{fig:pp}%
Current density as a function of the gap distance $d$. The current density for the TFA barrier (black dashed line) is compared to the LDA calculations (red $\times$) and two possible fits using GG barrier (red dashed lines).
}
\end{figure}

The current-density dependence with respect to the distance is shown in Fig.~\ref{fig:pp}(a).
The current for both TFA and LDA shows an exponential behavior with respect to $d$. Nevertheless, the difference between the two results is of about 3 order of magnitude. Two different fits for the LDA barrier are shown (red dashed lines) depending on the fit of the three parameters $u_0$, $\beta$ and $\alpha$ with a GG function, or just two (fixing $u_0$ to the height of the LDA barrier).

Similarly, in Fig.~\ref{fig:pp}(b) we represent the electronic heat flux for the TFA (black) and LDA barrier (in red, with two possible GG fits). The electronic heat fluxes (dashed lines) are compared with the total contribution (solid lines) that include the radiative heat transfer. Interestingly, this curve lead to rather different conclusions in terms of the distance below which electronic heat flux goes above the radiative one (more than 1\,nm for LDA, around 7\,\AA{} for TFA).

\subsection{Tip--plane configuration}

As explained above, while the experimental challenges associated with parallelism make the plane--plane scenario rather complicated to implement, the tip--plane configuration is much more convenient and widely used. In order to estimate the impact of barrier height and shape in this geometry, we exploit the Derjaguin or proximity force approximation (PFA)~\cite{Derjaguin68}, typically employed in different contexts (including but not limited to near-field radiative heat transfer) to deal with complex geometries by exploiting the results from the plane--plane configuration. Other geometry-dependent methods exist, but they are challenging to implement for small gap sizes due to slow numerical convergence~\cite{song}. For a spherically-shaped tip of radius $R$, the net power exchanged between the tip and the sample, in the absence of applied bias, can be written as
\begin{subequations}
\label{eq:PFA}
\begin{equation}
	P=P^{\rm (rad)}+P^{\rm (el)},
\end{equation}
in agreement with the net flux defined in Eq.~\eqref{eq:totalflux}, where each term is defined as
\begin{equation}
	P^{(Q)}=2\pi\int_0^R \mathrm{d} s \; s\;\Phi^{ (Q)}\!\left(d + R-\sqrt{R^2-s^2}\right),
\end{equation}
\end{subequations}
where $Q\in\{\mathrm{rad,el}\}$. The PFA calculation uses the results from the plane--plane configuration and considers the tip as a collection of rings at different distances from the plane, as illustrated in Fig.~\ref{fig:schema}(b). For electrons, almost all the tunneling heat comes from the tip apex as quantitatively shown in App.~\ref{app:PFA}.
As in Eq.~\eqref{eq:totalflux}, we also neglect the phonon tunneling contribution in these equations and consider rigid electrodes. The phononic contribution has been shown to be up to 10\% of the electronic flux for angstrom gaps when using a fluctuational approach and in the absence of bias~\cite{nottingham23} and at contact electrons are the main carrier attributed to the thermal conductivity of metals~\cite{ashcroft}. However the suitability of the PFA and the influence of geometry for phonons remains unexplored.

\begin{figure}[t]
\includegraphics[width=\linewidth]{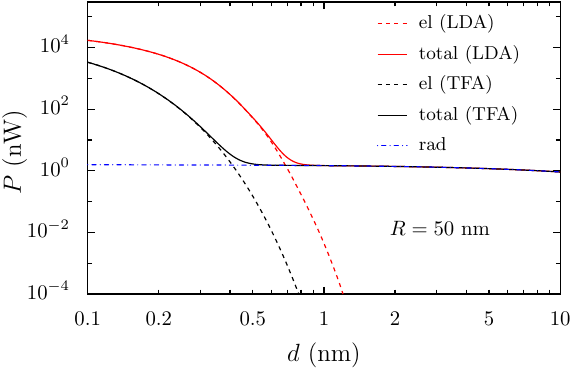}
\caption{\label{fig:pfa}%
Heat power emitted by a tip of radius $R=50$ nm, as a function of the distance $d$ for temperatures $T_1=400$ K and $T_2=300$ K. Two electronic contributions are shown based on the model of the TFA barrier (black dashed) and LDA one fitted with a GG distribution (red dashed). Solid lines include the radiative contribution (blue dash-dotted).
}
\end{figure}

In Fig.~\ref{fig:pfa}, we compare the different contributions to the heat emitted by a tip of radius $R=50$ nm.
Due to numerical integration, the distance where the electronic contribution (black and red dashed) dominate over the radiative contribution (blue dotdashed) is slightly shorter than in the plane--plane configuration (cf. Fig.~\ref{fig:pp}).
As the electronic flux is almost exponential, most of the contribution is being emitted from a small percentage of the tip apex, as expected. However to account for the radiative contribution to the emitted power, one must consider a circular region with a radius larger than half of the tip radius. We obtain that the difference between the TFA barrier (black dashed) and the LDA barrier (red dashed) can make this distance vary by up to half a nanometer, for the radius considered here.

As of current experiments, this increase in the heat flux in the extreme near field regime due to the electronic contribution is either not detectable as in Ref.~\cite{reddy_17} or larger contributions appear at larger distances as in Ref.~\cite{kittel_17}. For the latter case, the presence of contamination has sometimes been suggested~\cite{guo22} and the presence of a bias voltage could have an additional influence~\cite{nottingham23}.

The magnitude of the exchanged power $P$ obtained with the LDA approach at $4\,$\AA{} of separation distance (see Fig.~\ref{fig:pfa}), corresponding to the lattice constant of gold, is comparable with the one measured experimentally in Kittel’s scanning thermal microscopy experiment \cite{kittel_17} (although a problem related to the definition of physical contact still remains in this experiment where an unexpected increase of the heat flux is observed at nanometric separation distances). This power corresponds to an effective conductivity $\kappa_{\rm eff}=P/(T_2-T_1)d\simeq 10\,$W/m/K, which remains smaller than the thermal conductivity of metals ($\kappa_{\rm Au}=318\;$W/m/K at $T=300\;$K), a value which can be considered as an unsurpassable limit. On the contrary, using the TFA approach, the thermal conductivity is two orders of magnitude smaller. This tends to demonstrate that the use of the LDA leads to a transfer which is overestimated near the contact, since the presence of an external bias voltage will further increase the transfer. 
 
\section{Conclusions}
\label{sec:conclusions}

In this work, we have investigated the electronic current and associated heat flux between two parallel metallic slabs separated by a vacuum gap in the nanometer and sub-nanometer range of distances (extreme near field). We have first shown that both quantities strongly depend on the description of the electronic barrier in the gap. More specifically, we have compared an approach based on the solution to the nonlocal Poisson equation in the Thomas-Fermi approximation to the numerical solution of the Kohn-Sham equations in the local-density approximation for jellium. We have shown that these approaches lead to quite different effective electronic barrier potentials (both in shape and height), and that the resulting electronic current and heat flux may differ by several orders of magnitude. Besides, by employing a generalized Gaussian shape for the barrier, we have confirmed the extreme sensitivity of both quantities with respect to the distribution parameters, describing barrier shape, height and degree of penetration inside the metallic slabs. Also, our results based on an accurate $S$-matrix-scheme have confirmed the limits of the widely-employed semiclassical WKB approach to deal with the transmission probability of electrons through an arbitrary potential barrier. 

Moreover we have seen that while the LDA leads to an overestimated transfer near the contact, the TFA seems to be a more realistic approach since it allows reproducing the magnitude of heat flux measured in the recent experiments. However the presence of an external bias voltage has not been considered in the present study. It will require specific attention in a future work.

Our results show how quantitatively relevant is the choice of both the electronic barrier shape and the transmission-probability calculation scheme to obtain a reliable value of both current and heat flux. Apart from its fundamental interest, we have shown that in the context of extreme-near-field heat transfer this discrepancy can have an impact on the threshold distance at which electronic flux competes and goes beyond the radiative one. More generally, our results could be relevant for a more realistic modeling of experimental setups involving scanning thermal microscopy.

\begin{acknowledgments}
This research was supported by the French Agence Nationale de la Recherche (ANR), under grant ANR-20-CE05-0021-01 (NearHeat). 
\end{acknowledgments}
\appendix
\section{Computational details of LDA effective potential calculation}
\label{app:DFT}

For all LDA barrier calculations we use the grid-based projected augmented wave (GPAW) open source toolkit~\cite{gpaw1,gpaw2}. Each cell is composed of two jellium slabs of thickness equal to 4 times the lattice constant separated by a vacuum gap as in Fig.~\ref{fig:schema}. For all the calculations we consider a $4 \times 4 \times 4$ supercell with periodic boundary conditions, with a plane-wave cutoff energy of 400\,eV. The grid spacing starts at 0.2\,\AA{} and is reduced until finding a convergent barrier shape. The number of electronic bands in the calculation is set equal to the number of electrons in each cell, proportional to the volume of metal on each side.

For gold we use a lattice constant of 4.078 \AA{} and a Wigner-Seitz radius of 3.02 Bohr radii.
\section{Tip depth contributing to the emitted PFA power}
\label{app:PFA}	

\begin{figure}[t]
\includegraphics[width=\linewidth]{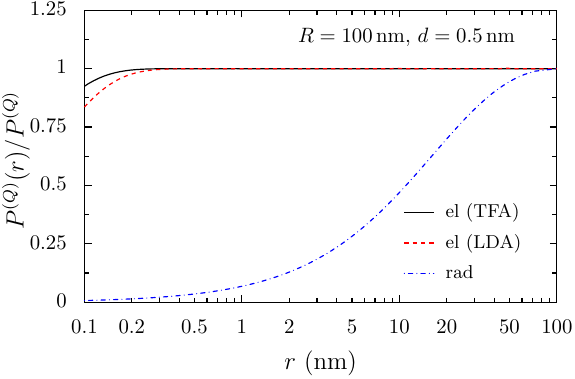}
\caption{\label{fig:app_pfa}%
Ratio between the partial power $P^{(Q)}(r)$ over the total power $P^{(Q)}$ per carrier $Q=\mathrm{el},\mathrm{rad}$, calculated using proximity force approximation, of the electronic (TFA black and LDA red) and radiative (blue) contributions as a function of tip depth $r$. The tip radius is $R=100$~nm and a gap of $d=5$~\AA{}.
}
\end{figure}

It is often considered in STM experiments only the last atom at the tip apex is responsible for the tunneling~\cite{chenSTM}. We can confirm that this behavior is also reproduced under the PFA. Due to the different power laws of the heat flux as a function of distance for the different carriers, their behavior is different under PFA. We define the partial power as
\begin{subequations}
\label{eq:PFAr}
\begin{equation}
	P(r)=P^{\rm (rad)}(r)+P^{\rm (el)}(r),
\end{equation}
in agreement with the net flux of Eq.~\eqref{eq:totalflux}, where each term is defined as
\begin{equation}
	P^{(Q)}(r)=2\pi\int_0^r \mathrm{d} s \; s\;\Phi^{ (Q)}\!\left(d + R-\sqrt{R^2-s^2}\right),
\end{equation}
\end{subequations}
where the equations are analogous to Eq.~\eqref{eq:PFA} but the integration goes from the tip apex up to a distance $r<R$.
In Fig.~\ref{fig:app_pfa}, we show the electronic and radiative contributions to $P(r)$ divided by the total power $P$, as a function of the tip depth $r$. 


\begin{thebibliography}{}
\bibitem{constants} Throughout this manuscript we use the constant values for gold for all the calculations, which can be found in Ref.~\cite{nottingham23}.
\end{thebibliography}
\end{document}